\documentclass[12pt, ams, onecolumn, amssymb, floatfix, longbibliography]{revtex4-2}

\usepackage{amsmath}
\usepackage{color}
\usepackage{amsfonts}
\usepackage{amssymb}
\usepackage{graphicx}
\usepackage{geometry}
\usepackage{hyperref}
\usepackage{comment}
\usepackage{tabularx}
\usepackage{bm}
\usepackage{euscript}
\usepackage{graphicx}
\usepackage{color}
\usepackage{amsfonts}
\usepackage{exscale}
\usepackage{amsbsy}
\usepackage{textcomp}
\usepackage{comment}
\usepackage{hyperref}
\usepackage{slashed}
\usepackage{mathtools}
\usepackage{tabularx}
\usepackage{bm}
\usepackage{euscript}
\usepackage{graphicx}
\usepackage{color}
\usepackage{amsfonts}
\usepackage{exscale}
\usepackage{amsbsy}
\usepackage{subfig}
\usepackage{textcomp}
\usepackage{comment}
\usepackage{hyperref}
\usepackage[dvipsnames]{xcolor}
\usepackage{natbib}

\newcommand{\jh}{\kappa}
\newcommand{\Ea}{m_{+}}
\newcommand{\Eb}{m_{-}}

\newcommand{\Tr}{\text{Tr}}

\newcommand{\pju}{P_{1}}
\newcommand{\pjd}{P_{2}}

 \newcommand{\tdj}{ \frac{\slashed{\nabla} }{i} }

 \newcommand{\ds}{\slashed{\nabla}}

\newcommand{\pd}{\partial}

\newcommand{\nn}{\nonumber}

\numberwithin{equation}{section}

\begin{document}

\title{Composite Fermion Nonlinear Sigma Models}
\author{Chao-Jung Lee}
\affiliation{Department of Physics, California Institute of Technology, Pasadena, CA 91125, USA}
\author{Prashant Kumar}
\affiliation{Department of Physics, Princeton University, Princeton, New Jersey 08544, USA}
\author{Michael Mulligan}
\affiliation{Department of Physics and Astronomy, University of California, Riverside, CA 92511, USA}

 \bigskip
 \bigskip
 \bigskip

\begin{abstract}
We study the integer quantum Hall plateau transition using composite fermion mean-field theory.
We show that the topological $\theta = \pi$ term in the associated nonlinear sigma model [P.~Kumar et al., Phys.~Rev.~B {\bf 100}, 235124 (2019)] is stable against a certain particle-hole symmetry violating perturbation, parameterized by the composite fermion effective mass.
This result, which applies to both the Halperin, Lee, and Read and Dirac composite fermion theories, represents an emergent particle-hole symmetry.
For a disorder ensemble without particle-hole symmetry, we find that $\theta$ can vary continuously within the diffusive regime.
Our results call for further study of the universality of the plateau transition.
\end{abstract}

\maketitle

\bigskip

\newpage

\tableofcontents

\newpage

\vskip 1cm

\section{Introduction}

Composite fermions \cite{Fradkinbook, jainCF} provide a powerful alternative perspective of the quantum Hall effect, one in which the integer and fractional effects are realized as integer quantum Hall mean-field states.
Over the last few years the possible relevance of composite fermion mean-field theory to the entire phase diagram and, in particular, quantum Hall plateau transitions has been emphasized.
Although mean-field theory is inadequate to fully account for the effects of electron interactions, semiclassical reasoning \cite{2017PhRvX...7c1029W} and numerical calculations \cite{2019PhRvB..99b4205Z, PhysRevB.98.115105, PhysRevLett.126.056802} indicate that composite fermion mean-field theory has several advantages over the noninteracting electron approach \cite{RevModPhys.67.357} (see, e.g., \cite{ZIRNBAUER2019458, PhysRevB.99.235114, 2020arXiv200809025S} for some recent work): 
(1) integer and fractional quantum Hall transitions \cite{SondhiGirvinCariniShahar} are united within a single composite fermion framework (e.g., \cite{ShimshoniSondhiShahar1997, PhysRevB.99.125135});
(2) composite fermion mean-field theory produces finite, {\it nonzero} quantum critical conductivity without recourse to residual electron interactions \cite{PhysRevB.61.8326};
(3) composite fermion theories can manifestly preserve \cite{Son2015, Seiberg:2016gmd, PhysRevB.99.205151} possible emergent reflection symmetries of the electron system \cite{PhysRevLett.124.156801}.

There has been significant progress towards an analytical description of this quantum criticality \cite{PhysRevB.100.235124, 2020arXiv200611862K}.
Specifically in \cite{PhysRevB.100.235124} it was shown that the nonlinear sigma model (NLSM) for composite fermion diffusion, 
\begin{align}
\label{NLSMgeneral}
S_{\rm NLSM} = 
\int d^2x\ {\rm Tr} \Big( {1 \over 2 g} (\partial_j Q)^2 + i {\theta \over 16 \pi} \epsilon_{ij} Q \partial_i Q \partial_j Q \Big),
\end{align}
contains a topological $\theta = \pi$ term.
Here, $Q \in U(2n)/U(n) \times U(n)$ parameterizes composite fermion charge density fluctuations and the replica $n \rightarrow 0$ limit is understood.
This result holds for both the Halperin, Lee, and Read (HLR) \cite{halperinleeread} and Dirac \cite{Son2015, WangSenthilfirst2015, MetlitskiVishwanath2016} composite fermion theories; it assumes a quenched disorder ensemble that preserves particle-hole symmetry (see below Eq.~\eqref{disorderensemble}).
In \eqref{NLSMgeneral} $g \propto 1/\sigma_{xx}^{\rm cf}$ is a marginally-relevant coupling \cite{Brezin:1980ms} that characterizes the evolution from the ballistic to the diffusive regime of finite (impurity-averaged) conductivity $\sigma^{\rm cf}_{ij}$, measured in units of $e^2/h$.
The topological $\theta$ term \footnote{The $\theta$ angle is well defined modulo $2\pi$ because ${1 \over 16 \pi} \int \epsilon_{ij} {\rm Tr} (Q \partial_i Q \partial_j Q)$ is integrally quantized for smooth configurations of $Q$.}---familiar from the seminal work of Pruisken et al.~\cite{PRUISKEN1984277, PhysRevLett.51.1915} on the integer quantum Hall transition and from theoretical studies of disordered Dirac fermions in graphene and related systems \cite{ALTLAND2002283, PhysRevB.66.165304, PhysRevLett.98.256801, PhysRevB.77.195430, PhysRevLett.97.236802}---is believed to prevent localization and thereby provide an explanation for the diffusive quantum criticality of the transition.
$\theta = \pi$ indicates particle-hole symmetric dc electrical transport \cite{kivelson1997, Son2015}: 
For the nonrelativistic (HLR) composite fermion, $\theta = 2\pi \sigma_{xy}^{\rm cf}$ (mod $2\pi$) \cite{PRUISKEN1984277}; for the Dirac composite fermion, $\theta = \pi + 2 \pi \sigma_{xy}^{\rm cf}$ (mod $2\pi$) \cite{PhysRevB.100.235124, PhysRevLett.98.256801}.

In this paper we consider the stability of this result to certain particle-hole symmetry violating perturbations.
Consider the mean-field composite fermion Lagrangians for HLR ($\eta = 1$) and Dirac ($\eta = 0$) composite fermion theories (see Appendix \ref{CFmeanfieldtheory} for a review), expressed in a standard Dirac notation,
\begin{align}
\label{meanfieldLagrangian}
{\cal L} = \bar \Psi \big(i \slashed{\partial} + \slashed{a} \big)\Psi - \eta \Psi^\dagger P_2 \big(i \partial_0 + a_0 \big)\Psi + m_1 \Psi^\dagger \Psi + m_2 \bar \Psi \Psi.
\end{align}
Above, $a_0$ and $a_j$ for $j \in \{1,2 \}$ are uncorrelated quenched scalar and vector potential disorders; $P_2$ projects onto the second component of $\Psi$; $m_1$ is a chemical potential determined by the composite fermion density; and $m_2$ is a mass that is odd under particle-hole symmetry.
We generalize the study in \cite{PhysRevB.100.235124} to include uniform $m_2$ and random $a_0$.
We show the $\theta = \pi$ term to be stable to the addition of nonzero $m_2$.
This represents an emergent particle-hole symmetry of the diffusive quantum critical point.
When $a_0$ is included, we find that the topological $\theta$ term varies continuously with the strength of this particle-hole symmetry violating disorder.
We are unable to determine within this NLSM approach (valid for $\sigma_{xx}^{\rm cf} \gg 1$) whether the ultimate low temperature fixed point of the sigma model with $\theta \neq \pi$ is an insulator or a quantum critical metal with varying Hall conductivity. 

The remainder of this paper is organized as follows.
We start with a description of the generating functional for disorder-averaged products of retarded and advanced composite fermion Green's functions.
Encoded in this generating functional are observables such as the composite fermion density of states and conductivity.
We then derive the NLSM for composite fermion mean-field theory, focusing on the topological $\theta$ term.
We separately consider the effects of particle-hole symmetry preserving and particle-hole breaking quenched disorder.
We conclude with a discussion of the HLR and Dirac composite fermion NLSMs and of possibilities for future work.
HLR and Dirac composite fermion mean-field theories are defined in Appendix \ref{CFmeanfieldtheory}.
Additional appendices supplement arguments in the main text.
Unless stated otherwise we take $e^2 = \hbar = 1$.

\section{Green's Function Generating Functional}

We begin with the generating functional of composite fermion Green's functions.
For a suitable choice of parameters this generating functional applies to both the HLR and Dirac composite fermion theories.
We discuss both, in turn, before focusing for definiteness on the HLR theory in the next section.

\subsection{Setup}
\label{setup}

\subsubsection{HLR}

Composite fermion mean-field states are described by a 2d field theory.
Within the replica approach (see, e.g., \cite{nayakmanybody, altlandsimonsbook}), disorder-averaged retarded and advanced HLR composite fermion Green's functions obtain from the disorder average of the path integral,
\begin{align}
Z = \int {\cal D}[\psi] {\cal D} [\psi^\dagger] e^{- S}
\end{align}
with action
\begin{align}
\label{starthlr}
S & = - \int d^2 x\ \psi^\dagger \Big( {1 \over 2m} D_j^2 + V + E_F + (\omega +  i \epsilon) \tau^3 \Big) \psi.
\end{align}
Here $\psi^\dagger = \psi^\dagger_{I}({\bf x})$ creates an HLR composite fermion of energy $E_F + \omega$ at ${\bf x} = (x_1, x_2)$; 
$I \in \{1, \ldots, n\}$ refer to retarded fermions and $I \in \{n+1, \ldots, 2n \}$ refer to advanced fermions; 
$D_j = \partial_j - i a_j$ is a covariant derivative with respect to possible quenched vector potential disorder $a_j({\bf x})$; 
$\epsilon > 0$ is an infinitesimal; and the diagonal matrix $\tau^3_{IJ} = \delta_{IJ}$ for $I, J \in \{1, \dots, n \}$ and $\tau^3_{IJ} = - \delta_{IJ}$ for $I, J \in \{n + 1, \dots, 2n \}$.
This last term reduces the symmetry of $S$ to $U_R(n) \times U_A(n) \subset U(2n)$, thereby distinguishing retarded and advanced fermions.
The replica limit $n \rightarrow 0$ is understood to be taken after disorder averaging any correlation function obtained from $Z$.

Within this mean-field approach $E_F > 0$ is the Fermi energy and $m/E_F$ is a finite nonzero parameter. 
In particular, we don't constrain the composite fermion mass $m$ by Kohn's theorem \cite{PhysRev.123.1242}. 
See Appendix \ref{hlrmeanfieldtheory} for additional details about HLR mean-field theory.

As a consequence of flux attachment, HLR mean-field theory realizes weak, quenched electron scalar potential disorder $V({\bf x})$ as anticorrelated vector and scalar potential randomness \cite{KimFurusakiWenLee1994, 2017PhRvX...7c1029W}:
\begin{align}
\label{slaving}
\epsilon_{i j} \partial_i a_j({\bf x}) = - 2 m V({\bf x}).
\end{align}
We take the vector potential to be a transverse Gaussian random variable with zero mean and variance $W > 0$:
\begin{align}
\label{disorderensemble}
\overline{a_i({\bf x})} = 0, \quad \overline{a_i({\bf x}) a_j({\bf x}')} = W \delta_{i j} \delta({\bf x} - {\bf x}').
\end{align}
Particle-hole symmetric disorder ensembles have vanishing odd moments $\overline{V({\bf x}_1) \cdots V({\bf x}_{2p + 1})} = 0$.
Delta-function correlated vector potentials \eqref{disorderensemble} arise from power-law correlated scalar potentials $\overline{V({\bf x}) V({\bf x}')} \propto |{\bf x} - {\bf x}'|^{-4}$ \cite{PhysRevB.95.075131}.
(Nonrelativistic fermions coupled to power-law correlated vector potential disorder (without scalar potential disorder) obtain singular single-particle properties; two-particle properties remain regular and appear to coincide with delta-function correlated vector potential disorder \cite{PhysRevLett.69.2979, PhysRevB.47.12051, PhysRevB.49.16609}.)

Following \cite{PhysRevB.100.235124} we relate $Z$ to a Euclidean 2d Dirac theory.
First we factorize the derivative terms in $S$,
\begin{align}
\label{startaction}
S = \int d^2x\ \Big(\psi^\dagger i v (D_1 + i D_2) \chi + \chi^\dagger i v (D_1 - i D_2) \psi - \psi^\dagger \big(E_F + (\omega + i \epsilon) \tau^3\big) \psi - 2 m v^2 \chi^\dagger \chi \Big),
\end{align}
using Eq.~\eqref{slaving} and the auxiliary fermion \cite{Son2015}:
\begin{align}
\label{auxrealrelations}
\chi = {i \over 2 m v} (D_1 - i D_2) \psi, \quad \chi^\dagger = - {i \over 2 m v} (D^\ast_1 + i D^\ast_2) \psi^\dagger
\end{align}
where the arbitrary velocity $v > 0$ can be set to unity upon a rescaling of the coordinates and vector potential disorder.
Next we define the Euclidean spinors \cite{1996PhLB..389...29V},
\begin{align}
\Psi = \begin{pmatrix} \psi \cr \chi \end{pmatrix}, \quad \Psi^\dagger = \begin{pmatrix} \psi^\dagger & \chi^\dagger \end{pmatrix},
\end{align}
and choose gamma matrices,
\begin{align} 
\gamma^1 = \begin{pmatrix} 0 & - i \cr i & 0 \end{pmatrix}, \quad \gamma^2 = \begin{pmatrix} 0 & 1 \cr 1 & 0 \end{pmatrix}, \quad \gamma^5 
= \begin{pmatrix} - 1 & 0 \cr 0 & 1 \end{pmatrix}.
\end{align}
$S$ becomes the Euclidean 2d Dirac action:
\begin{align}
\label{2dhlr}
S & = \int d^2x\ \Psi^\dagger \Big( \gamma^5 \gamma^j D_j - M_1 - M_2 \gamma^5  \Big) \Psi,
\end{align}
where the mass matrices are
\begin{align}
\label{masses}
M_1 = \big(m + {E_F \over 2}\big) \tau^0 + {\omega + i \epsilon \over 2} \tau^3, \quad M_2 = \big(m - {E_F \over 2} \big) \tau^0 + {\omega + i \epsilon \over 2} \tau^3,
\end{align}
where $\tau^0$ is the $2n \times 2n$ identity matrix.
We choose a regularization preserving the $U(2n)$ symmetry (present at $\omega = \epsilon = 0$) that identically rotates the spinor components $\psi = {1 \over 2} (1 - \gamma^5) \Psi$ and $\chi = {1 \over 2} (1 + \gamma^5) \Psi$. 

\subsubsection{Dirac}

The corresponding action that appears in the generating functional of Dirac composite fermion retarded and advanced Green's functions has precisely the same form as \eqref{2dhlr} with the replacement of the mass matrices by
\begin{align}
\label{diracmasses}
M_1^D = E_F \tau^0 + \big(\omega + i \epsilon \big) \tau^3, \quad M^D_2 = m_D \tau^0.
\end{align}
See Appendix \ref{diracmeanfieldtheory} for additional details.
At $\omega = \epsilon = 0$, the 2d actions governing HLR and Dirac Green's functions are identical upon the replacements $m + E_F/2 \leftrightarrow E_F$ and $m - E_F/2 \leftrightarrow m_D$.
Consequently, it's sufficient to consider the HLR generating functional for dc quantities and rename parameters as appropriate to find the corresponding result for Dirac composite fermion theory.

\subsection{Euclidean Discrete Symmetries}
\label{discretesymmetries}

Hermitian conjugation complex conjugates the masses and leaves the remaining terms unchanged:
\begin{align}
S^\dagger & = \int d^2x\ \Psi^\dagger \Big( \gamma^5 \gamma^j D_j - M_1^\ast - M_2^\ast \gamma^5  \Big) \Psi.
\end{align}
The retarded and advanced parts of $S$ are Hermitian conjugates at $\omega = 0$ for particle-hole symmetric disorder $a_j$.
We take charge conjugation to act as
\begin{align}
{\cal C} \Psi {\cal C} = \gamma^2 \Psi^\ast.
\end{align}
Since $\Psi^\dagger \gamma^5 \gamma^j \Psi$ and $\Psi^\dagger \Psi$ are odd under charge conjugation,
\begin{align}
\label{chargeeuclidean}
{\cal C} S {\cal C} & = \int d^2x\ \Psi^\dagger \Big( \gamma^5 \gamma^j D^\ast_j + M_1 - M_2 \gamma^5  \Big) \Psi.
\end{align}
This is consistent with the usual definition of charge conjugation (e.g., \cite{Peskin:1995ev}) and the identifications of $i M_1$ and $M_2$ as ``$\gamma_5$-type" and ``conventional" Dirac masses in the Minkowski signature version of this Euclidean action---see Appendix \ref{appendixminkowskiaction}.
Because Hermitian conjugation and Wick rotation don't commute, we use the Minkowski action \eqref{minkowskiaction} to determine that $\Psi$ and $\Psi^\dagger$ transform under a $U(1)$ chiral rotation as
\begin{align}
\label{chiraleuclidean}
\Psi \rightarrow e^{i \alpha \gamma^5} \Psi, \quad \Psi^\dagger \rightarrow \Psi^\dagger e^{i \alpha \gamma^5}.
\end{align}
The derivative terms in $S$ are invariant under continuous chiral transformations; the mass terms only preserve chiral transformations with $\alpha \in \pi \mathbb{Z}$.

The Dirac composite fermion mass $m_D$ changes sign under a particle-hole transformation  of the (2+1)d theory \cite{Son2015}.
Notice that $M_1$ is invariant under the combination of charge-conjugation and a chiral rotation with $\alpha = \pi/2$, while $M_2$ changes sign under this operation.
Consequently, we identify this transformation as the realization of electron particle-hole symmetry in the 2d Euclidean theory.
Consistent with this interpretation, we'll find that nonzero $M_2$ is associated to particle-hole symmetry violating Hall response.
 
\section{Effective Action for Charge Diffusion}
\label{sigmamodels}

In this section we derive the NLSM \eqref{NLSMgeneral} for HLR composite fermion mean-field theory dc charge diffusion.
This derivation holds for the Dirac composite fermion theory with the replacements: $m + E_F/2 \rightarrow E_F$ and $m - E_F/2 \rightarrow m_D$ (see \eqref{masses}, \eqref{diracmasses} and Appendix \ref{diracmeanfieldtheory}).
We focus on the $\theta$ angle and its physical interpretation; the derivation of the leading non-topological term in the NLSM and further details on the calculation of the topological term can be found in Appendix \ref{nlsmderivation}.

\subsection{Particle-hole Symmetric Disorder}
\label{phsymmetrysigmamodel}

In the dc zero-temperature limit we set the frequency $\omega = 0$ and use Eq.~\eqref{disorderensemble} to replace the action in \eqref{2dhlr} with its disorder average:
\begin{align}
\overline S = \overline S^{(1)} + \overline S^{(2)},
\end{align}
where
\begin{align}
\label{S1}
\overline S^{(1)} & = \int d^2x\ \Psi^\dagger \Big( \gamma^5 \gamma^j \partial_j - M_1 - M_2 \gamma^5  \Big) \Psi, \\
\label{S2}
\overline S^{(2)} & = {W \over 2} \int d^2x\ \big(\Psi^\dagger_{I} \gamma^5 \gamma^j \Psi_{I}\big) \big(\Psi^\dagger_{J} \gamma^5 \gamma^j \Psi_{J}\big),
\end{align}
and the parentheses around fermion bilinears indicate gamma matrix index contraction, e.g., $\big(\Psi^\dagger_{I} \gamma^5 \gamma^j \Psi_{I}\big) \equiv \Psi^\dagger_{a, I} \gamma^5_{ab} \gamma^j_{bc} \Psi_{c, I}$ for $a,b,c \in \{1,2\}$.
To understand the effects of the interaction
$\overline S^{(2)}$, we rewrite it as
\begin{align}
\overline S^{(2)} & = {W \over 2} \int d^2x\ \Big(\big(\Psi^\dagger_{I} \Psi_J \big) \big( \Psi_J^\dagger \Psi_I \big) - \big(\Psi^\dagger_{I} \gamma^5 \Psi_J \big) \big( \Psi_J^\dagger \gamma^5 \Psi_I \big) \Big)
\end{align}
and decouple the resulting 4-fermion terms using the Hubbard-Stratonovich fields $X_{IJ}$ and $Y_{IJ}$,
\begin{align}
e^{- \overline S^{(2)}} = \int {\cal D}[X]{\cal D}[Y] e^{- \int d^2 x \Big( {1 \over 2 W} {\rm Tr} X^2 + {1 \over 2 W} {\rm Tr} Y^2 - i X_{IJ} \Psi^\dagger_J \Psi_I - Y_{IJ} \Psi^\dagger_J \gamma^5 \Psi_I \Big)},
\end{align}
where ${\rm Tr} X^2 \equiv X_{JI} X_{IJ}$ and  ${\rm Tr} Y^2 \equiv Y_{JI} Y_{IJ}$.
$X_{IJ}$ and $Y_{IJ}$ each transform in the adjoint rep of $U(2n)$.

The saddle-point equations that determine $\langle X_{IJ} \rangle$ and $\langle Y_{IJ} \rangle$ have the form:
\begin{align}
\label{Xequation}
X & = {W \over 2} {\rm Tr}_\gamma \Big( \int {d^2 k \over (2\pi)^2} {i \over i \gamma^5 \gamma^j k_j - (M_1 + i X) - (M_2 + Y) \gamma^5} \Big), \\
\label{Yequation}
Y & = {W \over 2} {\rm Tr}_\gamma \Big( \int {d^2 k \over (2\pi)^2} {\gamma^5 \over i \gamma^5 \gamma^j k_j - (M_1 + i X) - (M_2 + Y) \gamma^5} \Big),
\end{align}
where the trace ${\rm Tr}_{\gamma}$ is only taken over the gamma matrix indices.
We consider an ansatz for $X_{IJ}$ and $Y_{IJ}$ that preserves the $U_R(n) \times U_A(n)$ symmetry:
\begin{align}
\label{PHsaddle}
\langle X \rangle = \Gamma \tau^3 + i \Sigma \tau^0, \quad \langle Y \rangle = Y_0 \tau^0,
\end{align}
where $\Gamma, \Sigma, Y_0$ are real.
We've ignored here and below a possible addition to the $\langle Y \rangle$ ansatz proportional to $\tau^3$ that appears to result in broken $U_R(n) \times U_A(n)$ symmetry.
$\Sigma$ is a logarithmic divergence that we absorb into a redefinition of the Fermi energy and composite fermion mass.
For general renormalized $M_2$ there's no solution to \eqref{Yequation} consistent with our ansatz; however when $M_2 = 0$ we find a nontrivial solution (at $\epsilon = 0$) indicative of composite fermion diffusion.
This solution produces the finite scattering rate $\Gamma \propto W E_F$ for {\it both} spinor components of $\Psi$.
This is different from the replacement of $\epsilon$ by $\Gamma$, which would instead only affect ${1 - \gamma^5 \over 2} \Psi$.
Eq.~\eqref{Yequation} is solved at finite $\Gamma$ by taking $Y_0 = 0$; thus, $Y_{IJ}$ is massive and can be ignored at sufficiently low energies.

The saddle-point solution requires vanishing particle-hole symmetry violating mass (recall \S\ref{discretesymmetries}).
Our interpretation is that even if initial conditions are chosen such that $M_2 \neq 0$, the saddle-point requires such particle-hole violating perturbations to renormalize to zero.
This implies the irrelevance of the $m_D$ mass in Dirac composite fermion mean-field theory near the diffusive integer quantum Hall transition; it represents an emergent particle-hole symmetry in the HLR theory.

The saddle-point solution spontaneously breaks $U(2n) \rightarrow U_R(n) \times U_A(n)$.
The Goldstone fluctuations of $X_{IJ}$ about this saddle-point are parameterized by writing $Q({\bf x}) \equiv X({\bf x})/\Gamma = U^\dagger({\bf x}) \tau^3 U({\bf x})$ where $U({\bf x}) \in U(2n)$.
This parameterization ensures that $Q({\bf x})$ satisfies $Q^2({\bf x}) = 1$.
Since $Q({\bf x}) = \tau^3$ for $U({\bf x}) \in U_R(n) \times U_A(n)$, the target manifold of $Q({\bf x})$ is $U(2n)/U_R(n) \times U_A(n)$.
The NLSM for $Q$ obtains by integrating out the fermions,
\begin{align}
\label{integrateoutaction}
e^{- S_{\rm NLSM}} = \int {\cal D} \Psi {\cal D} \Psi^\dagger e^{- \int d^2x\ \Psi^\dagger \Big( \gamma^5 \gamma^j \partial_j  - (M_1 + i \Gamma Q) - M_2 \gamma^5  \Big) \Psi}.
\end{align}
Since $M_2$ vanishes ($\epsilon = 0$ in the remainder) in the saddle-point solution, the derivation of $S_{\rm NLSM}$ from Eq.~\eqref{integrateoutaction} is identical to that in \cite{PhysRevB.100.235124}. 
In particular, $S_{\rm NLSM}$ contains a topological term \eqref{NLSMgeneral} with $\theta = \pi$.
(This is confirmed by the complementary analysis presented in the next section.)
The saddle-point solution that we found indicates that $\theta = \pi$ is stable against perturbation by particle-hole violating $M_2$.
This result holds for both the HLR and Dirac composite fermion theories

\subsection{Particle-hole Breaking Disorder}
\label{phbreakingsigmamodel}

Semiclassical reasoning and numerics \cite{2017PhRvX...7c1029W, PhysRevB.99.235114} suggest that HLR composite fermions exhibit a Hall conductivity in violation of particle-hole symmetry (which for HLR composite fermions requires \cite{kivelson1997} $\sigma^{\rm cf}_{xy} = - {1 \over 2}$ at $\sigma^{\rm cf}_{xx} < \infty$) if \eqref{slaving} isn't satisfied.
Here we explore the effects of this broken particle-hole symmetry within the NLSM approach.
(An alternative route to broken particle-hole symmetry that we don't pursue here is to consider a disorder ensemble with nonvanishing odd moments, in contrast to \eqref{disorderensemble}.)

To this end, we include additional scalar potential randomness $V_0(\mathbf{x})$ coupling to $\psi^\dagger \psi$ in \eqref{starthlr}---independent of the anticorrelated vector and scalar potential disorders in \eqref{slaving}---that has zero mean and variance $W_0$.
(Such a chiral coupling manifestly violates particle-hole symmetry in the Dirac composite fermion theory.)
The contribution of $V_0(\mathbf{x})$ to the disorder-averaged action $\overline S = \overline S^{(1)} + \overline S^{(2)} + \overline{S}^{(3)}$ is 
\begin{align}
\overline S^{(3)} = - {W_0 \over 2} \int d^2 x \big(\Psi_I^\dagger {1 - \gamma^5 \over 2} \Psi_I \big) \big(\Psi_J^\dagger {1 - \gamma^5 \over 2} \Psi_J \big),
\end{align}
where $\overline S^{(1)}$ and $\overline S^{(2)}$ are given in Eqs.~\eqref{S1} and \eqref{S2}.
In parallel with our treatment of $\overline S^{(2)}$ discussed in the previous section, we decouple the ``chiral disorder" in $\overline S^{(3)}$ with an additional Hubbard-Stratonovich field $Z_{IJ} \propto \Psi^\dagger_I {1 - \gamma^5 \over 2} \Psi_J$ and look for a saddle-point solution to the equations determining $\langle X_{IJ} \rangle$, $\langle Y_{IJ} \rangle$, and $\langle Z_{IJ} \rangle$:
\begin{align}
\label{Xequationnew}
X & = {W \over 2} {\rm Tr}_\gamma \Big( \int {d^2 k \over (2\pi)^2} {i \over i \gamma^5 \gamma^j k_j - (M_1 + i X + i {Z \over 2}) - (M_2 + Y - i {Z \over 2}) \gamma^5} \Big), \\
\label{Yequationnew}
Y & = {W \over 2} {\rm Tr}_\gamma \Big( \int {d^2 k \over (2\pi)^2} {\gamma^5 \over i \gamma^5 \gamma^j k_j - (M_1 + i X + i {Z \over 2}) - (M_2 + Y - i {Z \over 2}) \gamma^5} \Big), \\
\label{Zequation}
Z & = {W_0 \over 4} {\rm Tr}_\gamma \Big( \int {d^2 k \over (2\pi)^2} {1 - \gamma^5 \over i \gamma^5 \gamma^j k_j - (M_1 + i X + i {Z \over 2}) - (M_2 + Y - i {Z \over 2}) \gamma^5} \Big).
\end{align}
As detailed in Appendix \ref{XYZappendix}, we find a $U_R(n) \times U_A(n)$ preserving solution to these equations with the ansatz:
\begin{align}
\label{Xsolution}
\langle X \rangle & = \Gamma_2 \tau^3 + i X_0 \tau^0, \\
\label{Ysolution}
\langle Y \rangle & = Y_0 \tau^0, \\
\label{Zsolution}
\langle Z \rangle & = \big(\Gamma_1 - \Gamma_2 \big) \tau^3 + i Z_0 \tau^0,
\end{align}
where $\Gamma_{1,2} > 0$ and we've again ignored a possible term in the $\langle Y \rangle$ ansatz proportional to $\tau^3$.
We absorb the logarithmically divergent real constants $X_0$ and $Z_0$ into a renormalization of $m$ and $E_F$.
$Y_0$ is a finite real constant that allows for nonzero (renormalized) $M_2 = m - E_F/2 \propto W_0$.
The renormalized saddle-point mass matrix is therefore
\begin{align}
\label{massmatrix}
\begin{pmatrix} E_F \tau^0 + i \Gamma_1 \tau^3 & 0 \cr 0 & 2 m \tau^0 + i \Gamma_2 \tau^3 \end{pmatrix}.
\end{align}
Nonzero $W_0$ has produced unequal diffusion constants $\Gamma_{1}$ and $\Gamma_2$ for ${1 - \gamma^5 \over 2} \Psi$ and ${1 + \gamma^5 \over 2} \Psi$.

Similar to before, the saddle-point solution \eqref{Xsolution} - \eqref{Zsolution} spontaneously breaks the $U(2n)$ symmetry to $U(n)_R \times U(n)_A$.
We consider the fluctuations about this saddle-point by writing $Q_1({\bf x})  \equiv (X({\bf x}) + Z({\bf x}))/\Gamma_1 = U_1({\bf x}) \tau^3 U^\dagger_1({\bf x})$ and $Q_2({\bf x})  \equiv X({\bf x})/\Gamma_2 = U_2({\bf x}) \tau^3 U^\dagger_2({\bf x})$ where $U_1({\bf x}), U_2({\bf x}) \in U(2n)$. 
Since the action is only invariant under global ``vector" $U(2n)$ rotations (under which ${1 - \gamma^5 \over 2} \Psi$ and ${1 + \gamma^5 \over 2} \Psi$ transform identically) when $E_F$ or $m$ is nonzero, the Goldstone bosons correspond to those fluctuations of $Q_1$ and $Q_2$ for which $U_1 = U_2$; the ``axial" fluctuations $Q_1 \neq Q_2$ are massive (see Appendix \ref{axialmass} for an explicit demonstration) and can be neglected at low energies \footnote{The Goldstone phase of the $O(n)$ model perturbed by a $O(n-1)$ preserving field $h$ is a useful analogy here: at $h$ = 0 the Goldstone fluctuations are parameterized by $O(n)/O(n-1)$; at $h \neq 0$ the Goldstone fluctuations are parameterized by $O(n-1)/O(n-2)$ with the remaining degrees of freedom massive.}.
Thus we have a single light matrix boson $Q = Q_1 = Q_2 \in U(2n)/U_R(n) \times U_A(n)$ that we parameterize as $Q({\bf x}) = U({\bf x}) \tau^3 U^\dagger({\bf x})$ with $U({\bf x}) \in U(2n)$.

The NLSM for $Q$ obtains by integrating out the fermions.
The real part of $S_{\rm NLSM}$, which describes the diagonal conductivity of the system $1/g \propto W \gg 1$, is calculated in Appendix \ref{realapp}.
The imaginary part of $S_{\rm NLSM}$ is the topological $\theta$ term,
\begin{align}
S_{\rm top} =  i {\theta \over 16 \pi} \int d^2x\ {\rm Tr} \Big(\epsilon_{ij} Q \partial_i Q \partial_j Q \Big).
\end{align}
This term weights $Q$ configurations by the second homotopy group $\Pi_2\big(U(2n)/U_R(n) \times U_A(n) \big) = \mathbb{Z}$.
Since this classification is independent of the replica index $n$, we set $n=1$ and so $Q({\bf x}) = U({\bf x}) \tau^3 U^\dagger({\bf x}) \in U(2)/U_R(1) \times U_A(1) = SU(2)/U(1)$ with $U({\bf x}) \in SU(2)$.
To extract the topological term, it's convenient to perform the gauge transformation $\Psi \rightarrow U({\bf x}) \Psi$ before integrating out the fermions.
This allows us to interpret the fluctuations of $Q$ in terms of the $SU(2)$ gauge field,
\begin{align}
\label{gaugedef}
A_j = i U^\dagger \partial_j U.
\end{align}
Introducing the background fields,
\begin{align}
\label{backgroundfields}
\varphi_1 = m_+ \tau^0 + i \Gamma_+ \tau^3, \quad \varphi_2 = m_- \tau^0 + i \Gamma_- \tau^3,
\end{align}
for $m_\pm = m \pm E_F/2$ and $\Gamma_\pm = {\Gamma_2 \pm \Gamma_1 \over 2}$, we calculate $S_{\rm NLSM}$ to quadratic order in $A_j$,
\begin{align}
\label{integrateoutactionbreak}
e^{- S_{\rm NLSM}} = \int {\cal D}\Psi {\cal D} \Psi^\dagger e^{- \int d^2x\ \Psi^\dagger \Big( \gamma^5 \gamma^j (\partial_j - i A_j)  - \varphi_1 - \varphi_2  \gamma^5  \Big) \Psi}.
\end{align}
This is sufficient to obtain the topological term, which is cubic in $Q$.
Using \eqref{gaugedef} to relate $Q$ to $A_j$ and the identity $\epsilon_{jk} \partial_j A_k = i \epsilon_{j k} A_j A_k$ for a pure gauge potential, the topological $\theta$ term,
\begin{align}
\label{topgauge}
S_{\rm top} = {1\over 4 \pi} \int d^2x\ {\rm Tr} \Big(\tau^3 \epsilon_{j k} \big( \theta^{I} \partial_j A_k + i \theta^{II} A_j A_k \big) \Big)
\end{align}
and the $\theta$ angle,
\begin{align}
\theta = \theta^{I} + \theta^{II}.
\end{align}
Given the relation $\theta = 2 \pi \sigma^{\rm cf}_{xy}$, $\theta^{II}$ and $\theta^{I}$ are associated to the ``classical" and ``quantum" contributions to the Hall conductivity of the system \cite{Streda_1982}.

We now calculate each of these contributions to $\theta$.
Appendix \ref{nlsmderivation} contains additional details.
For particle-hole symmetry breaking $\varphi_2$---generated by $W_0$---we are unable to use the chiral anomaly argument in \cite{PhysRevB.100.235124} to determine $\theta$ \footnote{The issue is that a non-unitary chiral rotation is required to relate the retarded and advanced fermion mass matrices. We are unaware how the gauge field effective action changes under such to transformations.}.
Instead we combine a result of Goldstone and Wilczek \cite{PhysRevLett.47.986}, familiar from work on topological insulators \cite{qhz2008}, with a direct evaluation; a similar argument in this context can be found in \cite{PhysRevLett.98.256801}. 

Because $\theta^{I}$ in \eqref{topgauge} is only sensitive to the Abelian subgroup of $SU(2)$ generated by $\tau^3$, we can simplify its determination by focusing on the associated Abelian gauge field ${1 \over 2} {\rm Tr} (\tau^3 A_j)$ under which the retarded and advanced components of $\Psi$ carry opposite charge.
We'll furthermore treat $\varphi_{1}$ and $\varphi_2$ in \eqref{backgroundfields} as smoothly varying complex fields that assume their fixed values at the end of this calculation.
Writing $\varphi_{1,2}$ in terms of the complex fields $\chi^{R,A}_{1,2}$,
\begin{align}
\varphi_1 + \varphi_2 = \begin{pmatrix} e^{i \chi_1^R} & 0 \cr 0 & e^{i \chi_1^A} \end{pmatrix}, \quad \varphi_1 - \varphi_2 = - \begin{pmatrix} e^{- i \chi_2^R} & 0 \cr 0 & e^{- i \chi_2^A}   \end{pmatrix},
\end{align}
the generalization of \cite{PhysRevLett.47.986} to complex $\varphi_{1,2}$ gives the linear in $A_j$ contribution to $S_{\rm top}$:
\begin{align}
S_{\rm top}^{I} = {1 \over 8 \pi} \int d^2 x\ \epsilon_{j k} \partial_j \big(\chi^R_1 + \chi^R_2  - \chi^{A}_1 - \chi^{A}_2 \big) {\rm Tr} \Big( \tau^3 A_k \Big).
\end{align}
After an integration by parts, we identify
\begin{align}
\theta^I = - {1 \over 2} \Big(\chi^R_1 + \chi^R_2  - \chi^{A}_1 - \chi^{A}_2 \Big).
\end{align}

We now evaluate $\theta^I$ on the saddle-point solution in \eqref{backgroundfields}.
To ensure that $\theta^{I}$ is well defined modulo an integer multiple of $2\pi$, we constrain the real parts ${\rm Re}[\chi^{R,A}_{1,2}] \in [0, \pi)$.
For general $\varphi_{1,2}$ it's necessary to perform some combination of a charge conjugation \eqref{chargeeuclidean} and a chiral rotation with $\alpha = \pi/2$ \eqref{chiraleuclidean} on each fermion species in order to solve for $\chi^{R,A}_{1,2}$.
Note that a flip of the relative sign in $\chi^R_1 + \chi^R_2  - \chi^{A}_1 - \chi^{A}_2$ accompanies a charge conjugation on an advanced or retarded fermion.
Likewise any chiral rotation would generally include an additional contribution to the NLSM action due to the anomalous variation of the fermion measure \cite{PhysRevD.21.2848} in Eq.~\eqref{integrateoutactionbreak}.
For the $\varphi_{1,2}$ under consideration, it's only necessary to perform a charge conjugation on the advanced fermions with the result (see Appendix \ref{topone}):
\begin{align}
\label{exacttheta}
\theta^I = \pi + \arctan\big({\Gamma_+ - \Gamma_- \over m_+ - m_-}\big) - \arctan\big({\Gamma_+ + \Gamma_- \over m_+ + m_-} \big).
\end{align}
In the diffusive regime $0 < \Gamma_{1,2} \ll E_F$ with weak particle-hole symmetry violation $|\Gamma_1 - \Gamma_2| \propto W_0 \ll W$,  
\begin{align}
\label{approxtheta}
\theta^I = \pi + {\cal O}(W_0).
\end{align}

Eq.~\eqref{topgauge} indicates that $\theta^{II}$ is sensitive to the non-Abelian nature of $SU(2)$.
By direct evaluation of \eqref{integrateoutactionbreak} of the quadratic in $A_j$ contribution to $S_{\rm top}$, we find (see Appendix \ref{toptwo})
\begin{align}
\theta^{II} = {m_+ \Gamma_- - m_- \Gamma_+ \over m_+ \Gamma_+ - m_- \Gamma_-} \Big(\arctan\big({\Gamma_+ - \Gamma_- \over m_+ - m_-}\big) + \arctan\big({\Gamma_+ + \Gamma_- \over m_+ + m_-} \big) \Big).
\end{align}
 $\theta^{II} = 0$ when $W_0 = 0$.
Thus only $\theta^I$ contributes to $\theta$ when the disorder has particle-hole symmetry.

\section{Discussion}
\label{discussion}

In this note we showed the nonlinear sigma model of dc charge diffusion in HLR and Dirac composite fermion mean-field theories coincide in the presence of particle-hole symmetry preserving quenched disorder.
In particular, both the Dirac and HLR sigma models contain a $\theta = \pi$ term that is attractive with respect to a certain particle-hole violating perturbation (see Eq.~\eqref{meanfieldLagrangian}).
This topological term alters the renormalization group flow of the sigma model---which in its absence flows to a massive phase---towards a gapless fixed point.
Consequently, this topological term simultaneously serves to prevent localization and explain the diffusive quantum criticality of the integer Hall transition.
Our result shows how particle-hole symmetry can emerge in the HLR composite fermion theory and gives further evidence for the possible IR equivalence of the Dirac and HLR theories.

We have also showed how electron particle-hole breaking disorder can shift the $\theta$ angle away from $\pi$.
Since $\theta = 2\pi \sigma_{xy}^{\rm cf}$ (mod $2\pi$) for the HLR theory and $\theta = \pi + 2\pi \sigma_{xy}^{\rm cf}$ (mod $2 \pi$) for the Dirac theory, nonzero $W_0$ results in a violation of particle-hole symmetric electrical transport \cite{kivelson1997, Son2015}.
Because the nonlinear sigma model description is only appropriate for longitudinal conductivities $\sigma_{xx}^{\rm cf} \gg 1$ (in units of $e^2/h = 1$), we aren't able to determine the identity of the state that obtains for $\sigma_{xx}^{\rm cf} \leq 1$ towards which the nonlinear sigma model evolves.
(Recall that the longitudinal conductivity is a coupling in the nonlinear sigma model that runs towards zero.)
It would be interesting to understand if particle-hole symmetry emerges for $\sigma_{xx} \sim 1$, as predicted in \cite{PhysRevB.32.2636} and found in a numerical study of noninteracting electrons \cite{PhysRevLett.70.481}.
Alternatively, if particle-hole symmetry doesn't emerge, we expect either a gapped insulator or a diffusive metal (at least in the vicinity of the particle-hole symmetric limit at $\theta = \pi$).
It would be interesting to connect our result with recent work \cite{PhysRevLett.126.056802} that found evidence for a line of extended states with continuously varying exponents as particle-hole symmetry is violated by varying the coefficient of electron scalar potential $V({\bf x})$ in \eqref{slaving}.

\section*{Acknowledgments}

We thank Sri Raghu for useful conversations and correspondence.
This material is based upon work supported by the U.S. Department of Energy, Office of Science, Office of Basic Energy Sciences under Award Number DE-SC0020007. 
M.M. acknowledges the generous hospitality of the Stanford Institute for Theoretical Physics and support provided by the Moore Foundation for this work.
P.K. is supported by DOE-BES Grant Number DE-SC0002140.

\appendix

\section{Composite Fermion Mean-Field Theory}
\label{CFmeanfieldtheory}

In this appendix we derive the composite fermion mean-field theory Lagrangian \eqref{meanfieldLagrangian}.
For a suitable choice of parameters, this Lagrangian describes both the HLR and Dirac composite fermion theories and it forms the basis for our current study.

\subsection{HLR Mean-Field Theory}
\label{hlrmeanfieldtheory}

The HLR Lagrangian is \cite{halperinleeread}
\begin{align}
{\cal L}_{\rm HLR} = \psi^\dagger \Big(i \partial_0 + A_0 + a_0 - {\big(i \partial_j + A_j + a_j\big)^2  \over 2 m} \Big) \psi + {\epsilon^{\mu \nu \rho} \over 8 \pi} a_\mu \partial_\nu a_\rho + \ldots.
\end{align}
Here, $\psi^\dagger$ creates an HLR composite fermion; $A_\mu$ is the external electromagnetic field; $a_\mu$ is a dynamical (2+1)d Chern-Simons gauge field; $m$ is an effective mass, $\epsilon^{\mu \nu \rho}$ with $\mu, \nu, \rho \in \{0, 1, 2 \}$ is the antisymmetric symbol with $\epsilon^{012} = 1$; and the ``$\ldots$" include all other possible couplings and interactions, which we set to zero in the remainder of this appendix.
Variation of ${\cal L}$ with respect to $A_0$ implies the electron and composite fermion densities are equal.
A nonzero uniform magnetic field $\epsilon_{ij} \partial_i A_j = B > 0$ is assumed such that the electron filling fraction $\nu = 1/2$.

To write this Lagrangian in a Dirac form, we factorize the spatial derivative terms using the auxiliary fermion $\chi$ defined in \eqref{auxrealrelations}:
\begin{align}
\label{linearizedhlr}
{\cal L}_{\rm HLR} = \psi^\dagger \big(i \partial_0 + A_0 + a_0 \big) \psi - \psi^\dagger i v (D_1 + i D_2) \chi - \chi^\dagger i v (D_1 - i D_2) \psi + 2 m v^2 \chi^\dagger \chi + {\epsilon^{\mu \nu \rho} \over 8 \pi} a_\mu \partial_\nu a_\rho
\end{align}
where the arbitrary velocity $v > 0$.
(In the main text we rescale the spatial coordinates and the vector field to set $v= 1$ in the 2d theory associated to ${\cal L}_{\rm hlr}$.
The Fermi energy and frequency are then redefined to absorb $v$: $E_F/v^2 \rightarrow E_F$ and $\omega/v^2 \rightarrow \omega$.)
Notice that the scalar potential $A_0 + a_0$ only couples to $\psi$.

The mean-field approximation consists of imposing the $a_0$ equation of motion (i.e., flux attachment),
\begin{align}
\psi^\dagger \psi = - {1 \over 4 \pi} \epsilon^{ij} \partial_i a_j,
\end{align}
and then setting all dynamical fluctuations of $a_\mu$ to zero.
Introducing $\Psi = \begin{pmatrix} \psi & \chi \end{pmatrix}^T$ and the gamma matrices $(\Gamma^0, \Gamma^1, \Gamma^2) = (\sigma^3, i \sigma^1, i \sigma^2)$ for the Pauli-$\sigma$ matrices $\sigma^j$, the resulting mean-field Lagrangian is
\begin{align}
\label{hlrmflag}
{\cal L} = \bar \Psi \big(i \slashed{\partial} + \slashed{a} \big)\Psi - \Psi^\dagger P_2 \big(i \partial_0 + a_0 \big)\Psi + \big(mv^2  + {E_F \over 2}\big) \Psi^\dagger \Psi - \big(mv^2  - {E_F \over 2}\big) \bar \Psi \Psi,
\end{align}
where $\bar \Psi = \Psi^\dagger \Gamma^0$, $i \slashed{\partial} + \slashed{a} = i \Gamma^0 (\partial_0 + a_0) + i v \Gamma^j (\partial_j + a_j)$, $P_2 = \begin{pmatrix} 0 & 0 \cr 0 & 1 \end{pmatrix}$ projects onto the second component of $\Psi$, we've replaced $A_\mu + a_\mu$ by $a_\mu$, and the Fermi energy $E_F > 0$ fixes $\psi^\dagger \psi = B/4\pi$ on average.

\subsection{Dirac Mean-Field Theory}
\label{diracmeanfieldtheory}

The Dirac composite fermion Lagrangian is \cite{Son2015, WangSenthilfirst2015, MetlitskiVishwanath2016, Seiberg:2016gmd}
\begin{align}
{\cal L}_{\rm D} = \bar \Psi \big(i \slashed{\partial} + \slashed{a} \big)\Psi + m_D v^2 \bar \Psi \Psi - {1 \over 4 \pi} \epsilon^{\mu \nu \rho} A_\mu \partial_\nu a_\rho + {1 \over 8 \pi} \epsilon^{\mu \nu \rho} A_\mu \partial_\nu A_\rho,
\end{align}
where $\Psi$ is a 2-component Dirac fermion; $a_\mu$ is a dynamical (2+1)d gauge field; $A_\mu$ is the external electromagnetic field; $m_D$ is a $(2+1)$d Dirac mass; and the remaining terms are defined as below \eqref{hlrmflag}.
A uniform magnetic field $\epsilon_{ij} \partial_i A_j = B > 0$ and an electron filling fraction $\nu = 1/2$ are assumed.
In contrast to the HLR composite fermion theory, the scalar potential couples to both components of $\Psi$ in the Dirac theory.
Particle-hole symmetry is a manifest invariance of ${\cal L}_{\rm D}$ \cite{Son2015, Seiberg:2016gmd} at $m_D = 0$.

Similar to before, the mean-field approximation consists of imposing the $a_0$ equation of motion,
\begin{align}
\label{diracconstraint}
\Psi^\dagger \Psi = {B \over 4 \pi},
\end{align}
and setting all dynamical fluctuations of $a_\mu$ to zero.
The resulting mean-field Lagrangian is
\begin{align}
\label{diracmflag}
{\cal L} = \bar \Psi \big(i \slashed{\partial} + \slashed{a} \big)\Psi + E_F \Psi^\dagger \Psi + m_D v^2 \bar \Psi \Psi,
\end{align}
where $E_F > 0$ fixes \eqref{diracconstraint} on average.
Possible quenched randomness in $A_0$ sources random $a_j$ through the ${1 \over 4 \pi} \epsilon^{i j} A_0 \partial_{i j} a_j$ coupling in ${\cal L}_D$.

\section{Minkowski Action}
\label{appendixminkowskiaction}

In this appendix we translate the 2d Euclidean action $S$ in \eqref{2dhlr} into Minkowski signature.
Introduce the Minkowski spinors \cite{1996PhLB..389...29V},
\begin{align}
\label{MinkowskiEuclideanspinorrelation}
\Psi_M = e^{{\pi \over 4} \gamma^2 \gamma^5} \Psi, \quad \Psi_M^\dagger = \Psi^\dagger e^{{\pi \over 4} \gamma^2 \gamma^5},
\end{align}
(note the same transformation is used for $\Psi_M$ and $\Psi_M^\dagger$) and gamma matrices,
\begin{align}
\tilde \gamma^1 = \gamma^1, \quad \tilde \gamma^2 = - \gamma^5, \quad \tilde \gamma^5 = \gamma^2.
\end{align}
The action $S$ becomes
\begin{align}
- \int d^2x\ \Psi^\dagger_M \Big(\tilde \gamma^2 \tilde \gamma^j D_j - M_1 \tilde \gamma^2 \tilde \gamma^5 - M_2 \tilde \gamma^2 \Big) \Psi_M. 
\end{align}
Next we Wick rotate to Minkowski signature $(+1, - 1)$ by defining $x^0 = - i x^2, x^1 = x^1$ and $\gamma_M^0 = \tilde \gamma^2, \gamma_M^1 = i \tilde \gamma^1, \gamma_M^5 = i \tilde \gamma^5$.
The resulting Minkowski-signature action is 
\begin{align}
\label{minkowskiaction}
S_M & = i \int dx^0 dx^1 \bar \Psi_M \Big(i \gamma^\mu_M D_\mu - M_1 \gamma_M^5 - M_2 \Big) \Psi_M,
\end{align}
where $\mu \in \{0, 1 \}$, $\bar \Psi = \Psi_M^\dagger \gamma_M^0$, $D_\mu = \partial_\mu - i A_\mu$, and $A^0 = i a^2, A^1 = a^1$.
This action contains a $\gamma_5$-type Dirac mass (matrix) $i M_1$ and a conventional Dirac mass (matrix) $M_2$.
The chiral transformation in Minkowski signature is
\begin{align}
\label{chiralminkowski}
\Psi_M \rightarrow e^{i \alpha \gamma^5_M} \Psi_M, \quad \Psi^\dagger_M \rightarrow \Psi_M^\dagger e^{- i \alpha \gamma_M^5}.
\end{align}

\section{XYZ Saddle-Point Analysis}
\label{XYZappendix}

In this appendix we detail our saddle-point solution for particle-hole breaking disorder; our analysis of particle-hole preserving disorder obtains by taking $W_0 = 0$.
We begin with the saddle-point equations,
\begin{align}
\label{appXequationnew}
X & = {W \over 2} {\rm Tr}_\gamma \Big( \int {d^2 k \over (2\pi)^2} {i \over i \gamma^5 \gamma^j k_j - (m_+ + i X + i {Z \over 2}) - (m_- + Y - i {Z \over 2}) \gamma^5} \Big), \\
\label{appYequationnew}
Y & = {W \over 2} {\rm Tr}_\gamma \Big( \int {d^2 k \over (2\pi)^2} {\gamma^5 \over i \gamma^5 \gamma^j k_j - (m_+ + i X + i {Z \over 2}) - (m_- + Y - i {Z \over 2}) \gamma^5} \Big), \\
\label{appZequation}
Z & = {W_0 \over 4} {\rm Tr}_\gamma \Big( \int {d^2 k \over (2\pi)^2} {1 - \gamma^5 \over i \gamma^5 \gamma^j k_j - (m_+ + i X + i {Z \over 2}) - (m_- + Y - i {Z \over 2}) \gamma^5} \Big),
\end{align}
and the $U_R(n) \times U_A(n)$ ansatz,
\begin{align}
\label{appXsolution}
\langle X \rangle & = (\Gamma_+ + \Gamma_-) \tau^3 + i X_0 \tau^0, \\
\label{appYsolution}
\langle Y \rangle & = Y_0 \tau^0, \\
\label{appZsolution}
\langle Z \rangle & = - 2 \Gamma_- \tau^3 + i Z_0 \tau^0,
\end{align}
where $m_\pm = m \pm E_F/2$ and $\Gamma_\pm = {\Gamma_2 \pm \Gamma_1 \over 2}$.

We define the real parameters $J_1$ and $J_2$ by
\begin{eqnarray}
i J_1 \tau^3 + J_2 \tau^0 \equiv \int {d^2 k \over (2\pi)^2} {1 \over k^2 + (m_+ + i X + i {Z \over 2})^2 + (m_- + Y - i {Z \over 2})^2},
\end{eqnarray}
where by direct evaluation we find
\begin{align}
J_1 & = {{\rm sign} (\Gamma_- \tilde m_- - \Gamma_+ \tilde m_+ ) \over 8 \pi} \Big(\pi - 2 \arctan \Big({\tilde m_-^2 - \tilde m_+^2 - \Gamma_-^2  + \Gamma_+^2 \over 2 |\Gamma_- \tilde m_- - \Gamma_+ \tilde m_+|} \Big) \Big), \\
J_2 & = - {1 \over 8 \pi} \log\Big({\Lambda^4 \over \Big((\Gamma_+ + \Gamma_-)^2 + (\tilde m_+ + \tilde m_-)^2 \Big) \Big((\Gamma_+ - \Gamma_-)^2 + (\tilde m_+ - \tilde m_-)^2 \Big)} \Big),
\end{align}
with shifted
\begin{align}
\tilde m_+ & = m_+ - X_0 - Z_0, \\
\tilde m_- & = m_- - Y_0,
\end{align}
and UV cutoff $\Lambda$.
Written in terms of these parameters the saddle-point equations Eqs.~\eqref{appXequationnew} - \eqref{appZequation} become
\begin{align}
2 \Gamma_-
& = W_0  \, 
\big(J_1(\tilde m_+ + \tilde m_-) + J_2 (\Gamma_+ + \Gamma_-) \big), \\
 Z_0
& = - W_0 \; 
\big(J_1 (\Gamma_+ + \Gamma_-) - J_2 (\tilde m_+ + \tilde m_-) \big), \\
 \Gamma_+ + \Gamma_-
 & = - W  \,
 \big( J_1 \tilde m_+ + J_2 \Gamma_+ \big), \\
 X_0 
& =
- W \big(J_1  \Gamma_+ - 
 J_2 \tilde m_+ 
\big), \\
0 & = 
W 
\big(J_1 \tilde m_- + J_2 \, \Gamma_- \big), \\
Y_0
& =
- W 
\big( J_1 \, \Gamma_- + J_2 \tilde m_- \big).
\end{align}
We solve these equations for $X_0, Y_0, Z_0, \Gamma_\pm$, and $m_-$:
\begin{eqnarray}
&&  \label{sd-pt-sol-start}
m_- = \frac{ - m_+ \, W_0 \big( J_2 + W (J_1^2 + J_2^2) \big) }{2 + J_2 (3 W_0 + 4 W) + W (3 W_0 +2 W) \;(J_1^2 + J_2^2)}, \\
&&
Y_0 =
\frac{ - m_+ W_0 W (J_1^2+J^2_2)}{2 + J_2 (3 W_0 + 4 W) + W (3 W_0 +2 W) \;(J_1^2 + J_2^2)},  \\
&&
\Gamma_- = \frac{ E_1 W_0 J_1}{2 + J_2 (3 W_0 + 4 W) + W (3 W_0 +2 W) \;(J_1^2 + J_2^2)}, \\
&&
\Gamma_+ + \Gamma_- = \frac{ -2 m_+ W J_1}{2 + J_2 (3 W_0 + 4 W) + W (3 W_0 +2 W) \;(J_1^2 + J_2^2)},  \\
&&
Z_0 = \frac{2 m_+ W_0 \big( J_2 + W (J_1^2+J^2_2) \big)}{2 + J_2 (3 W_0 + 4 W) + W (3 W_0 +2 W) \;(J_1^2 + J_2^2)},  \\
&&
X_0  =\frac{m_+ W \big( 2J_2+ \big(W_0+ 2 W )(J_1^2+J^2_2) \big)}{2 + J_2 (3 W_0 + 4 W) + W (3 W_0 +2 W) \;(J_1^2 + J_2^2)}.
\label{sd-pt-sol-end}
\end{eqnarray}
Notice that $Y_0, \Gamma_-, Z_0$, and $m_-$ vanish as $W_0 \rightarrow 0$. 
We treat $m_- = m - E_F/2$ as a variable in order to avoid an overly constrained set of equations; 
our interpretation is that $W_0$ and $W$ disorders cause $m/E_F$ to renormalize to a value determined by the saddle-point equations.
To ensure the scattering rates $\Gamma_+ \pm \Gamma_- > 0$ of ${1 \pm \gamma^5 \over 2} \Psi$ are positive, we require 
\begin{align}
\label{positivecondition}
J_1 \Big(2 + J_2 (3 W_0 + 4 W) + W (3 W_0 +2 W) \;(J_1^2 + J_2^2 \Big) < 0.
\end{align}.

The retarded/advanced 2d Euclidean fermion Green's function that obtains from this saddle-point solution,
\begin{align}
\overline g_{R/A} = \langle \mathbf{x} | {1 \over i \sigma_1 \partial_1 - i \sigma_2 \partial_2 + \big(m_+ - X_0 - Z_0 + \omega \pm i \Gamma_+ \big) - \big(m_- - Y_0 - \omega \pm i \Gamma_- \big)\sigma_3}  | \mathbf{x}' \rangle
\end{align}
where $\omega$ parameterizes deviations about the Fermi energy, suggests how to absorb the logarithmic divergence $J_2$ into renormalized parameters.
We first introduce the renormalization factors,
\begin{align}
Z_R & \equiv  \frac{ 2 + J_2 (3 W_0 + 4 W)
+  W (3 W_0 +2 W) \;(J_1^2 + J_2^2)  }{ ( 2 + W_0\, J_2  + 2 W J_2 )  },\\
Z_L & \equiv \frac{  J_2  +   W (J_1^2+J^2_2)   }{ J_2 }.
\end{align}
Then the renormalized $m_\pm$ and $\Gamma_\pm$ are
\begin{align}
m_+^R & = {m_+ - X_0 - Z_0 \over Z_R}, \\
m_-^R & = {m_- - Y_0 \over Z_L}, \\
\Gamma^R_- & =  W_0  \; \frac{m_-^R    J_1   }{2 + (W_0 + 2 W) J_2},
 \\
\Gamma^R_+ & = - \big(2 W + W_0 \big)  \; \frac{m_-^R    J_1   }{2 + (W_0 + 2 W) J_2}.
\end{align}
The condition \eqref{positivecondition} ensures the renormalized $\Gamma^R_+ \pm \Gamma^R_-$ are positive.
We use these renormalized parameters (without the $R$ superscript) in the main text to find the $\theta$ angle.

\section{Detailed NLSM Derivation}
\label{nlsmderivation}

In this Appendix we detail the calculation of the NLSM for $Q$ that is sketched in \S \ref{phbreakingsigmamodel}.

\subsection{Setup}

Here we have found it convenient to employ a different convention than the one used in the main text.
The ``translation table" between the two conventions is given below.
\begin{eqnarray}
&&
\Gamma_1 = \frac{\alpha_2 + \alpha_1}{2}  \\
&& 
\Gamma_2 = \frac{\alpha_2 - \alpha_1}{2}   \\
&&
\Gamma_+ =\frac{\Gamma_2 + \Gamma_1}{2}
= \frac{\alpha_2}{2}  \\
&&
\Gamma_- =\frac{\Gamma_2 - \Gamma_1}{2}
= \frac{ - \alpha_1}{2}  \\
&& 
\gamma_1  = \sigma_y ,\;\;
\gamma_2 = \sigma_x , \;\;
\gamma_5 =  - \sigma_z.  
\end{eqnarray}
Above $(\sigma_x, \sigma_y, \sigma_z)$ are the usual Pauli sigma matrices.
Furthermore, we reflect both coordinates $(x, y) \rightarrow (-x, - y)$ and reverse the overall signs of the fermion and NLSM actions in \eqref{integrateoutactionbreak}:
\begin{align}
\label{appendixnlsm}
e^{S_{\rm NLSM}} = \int {\cal D} \Psi {\cal D} \Psi^\dagger\ e^{S},
\end{align}
with 
\begin{eqnarray}
S
=  \int d^2x\ \Psi^\dagger 
\Big(  i \sigma_x \partial_x 
 - i \sigma_y \partial_y 
  - \Eb \sigma_z 
+  \Ea   \sigma_0
+ i \frac{ \alpha_1 \sigma_z + \alpha_2 \sigma_0  }{2}  Q    
    \Big) \Psi.
\end{eqnarray}
Here $\sigma_0$ is the $2 \times 2$ identity matrix.
In the dc limit of interest, $m_\pm = m \pm E_F/2$.
The matrix boson $Q({\bf x}) = U({\bf x}) \tau^3 U^\dagger({\bf x})$ for $U({\bf x}) \in U(2n)$ where $\tau^3 = \sigma_z \otimes \sigma_0^n$.
Since all terms except for $Q$ are singlets with respect to the replica indices (i.e., the $\tau$ space), we (generally) leave implicit the $2n \times 2n$ identity matrix $\tau_0$ in this subspace.

We partially follow \cite{PhysRevLett.98.256801} to separately derive the real and imaginary parts of $S_{\rm NLSM}$.
Before proceeding, we define the self-consistent Born approximation (SCBA) Green's functions,
\begin{eqnarray}
g_{ \pm } \equiv  
\frac{1}{ 
 \Ea  \sigma_0
  +
 i \sigma_x \partial_x 
 - i \sigma_y \partial_y 
  - \Eb  \sigma_z 
\pm  i (\frac{ \alpha_1 \sigma_z + \alpha_2 \sigma_0  }{2})  \tau^3   
},
\label{gpm-def}
\end{eqnarray}
which are related to the retarded and advanced Green's functions $g_{R,A}$ as follows:
\begin{eqnarray}
\label{g-identity}
(g_+)^\dagger = g_-  \; ,  \;\; \; 
g_{+}= {\rm Diag}(g_R,g_A)_\tau   , \;\; \; 
g_{-}= {\rm Diag}(g_A,g_R)_\tau     \\
g_+   -  g_-   = (g_R-g_A) \, \tau^3   ,\;\; \; 
  g_+   +   g_-   = (g_R+g_A) \, \tau_0 . \;\;\; \;   
\end{eqnarray}
Note that the $\pm$ subscript of $g_\pm$ labels the sign of the imaginary part of the Green's function; it is unrelated to the $\pm$ subscript of the masses $m_\pm$.

\subsection{Real Part of $S_{\rm NLSM}$}
\label{realapp}

To compute the real part of $S_{\rm NLSM}$ we directly compute the fermion determinant implied by \eqref{appendixnlsm}:
\begin{align}
{\rm Re} \big(S_{\rm NLSM} \big) + i {\rm Im}\big(S_{\rm NLSM} \big) = \Tr \ln\Big( 
\big(  
  \Ea  \, \sigma_0
  - \Eb \, \sigma_z
+ i \sigma_x \pd_x - i \sigma_y \pd_y     \big) 
+    i \, \frac{ \alpha_1 \sigma_z  
+  \alpha_2  \sigma_0 }{2}     \,  Q  	   \Big).
\end{align}
For any operator $\hat X$ with determinant $\det \hat X = R e^{i \theta}$, we seek $\ln R$.
Let $i \hat{C} \equiv 
   i \, (\frac{ \alpha_1 \sigma_z  
+  \alpha_2  \sigma_0   }{2}   )  \,  Q  
$,    
$   \hat{B} \equiv  
\Ea  \, \sigma_0
  -  \Eb \, \sigma_z
+ i \sigma \pd_x - i \sigma_y \pd_y  $, 
we can decompose the operator in $2\times 2$ $\sigma$-space using the identities:
\begin{eqnarray}
&&
\det[ i \hat{C} + \hat{B} ]   
=    \det[ i \, \hat{C}]  + \det[\hat{B}] +i \,  \det[ \hat{C}] 
 \Tr[  \hat{C}^{-1} \,\hat{B} ], \\
&&
\det[ (\hat{B}+ i \hat{C}) \,\hat{H}^{-1} \, (\hat{B}-i \hat{C}) \, \hat{H} ]
=\Big{|}  \det[ i \hat{C} + \hat{B} ] \Big{|}^2 ,
\end{eqnarray}
where $\hat{H}$ is any constant matrix in $\sigma$-space that satifies $[\hat{C}, \hat{H}] =0$.
We choose $\hat{H} = (\frac{ \alpha_1 \sigma_z  
+  \alpha_2  \sigma_0   }{2}   ) $.
Up to an unimportant constant that we drop, we find
\begin{eqnarray}
{\rm Re} \big(S_{\rm NLSM} \big) = &&
\frac{1}{2} \Tr \,\ln
\Big[ (g^{-1}_+   - i \, \hat{H}   \tau^3
  + i \, \hat{H}\, Q  ) \, (\hat{H}  )^{-1}
  \,   (g^{-1}_-   +   i \, \hat{H} \, \tau^3 -  i\, \hat{H} \,  Q  )  
\Big]  \cr
&&
=
 \frac{1}{2} \Tr\Big[
 \big(   \frac{i}{2}(g_R-g_A) \,  \ds Q   \big)  
 -\frac{1}{2} \big(   \frac{i}{2}(g_R-g_A) \,  \ds Q   \big)     
   \,\big(   \frac{i}{2}(g_R-g_A) \,  \ds Q   \big)   
\Big]  \cr
&&
=
\frac{1}{16}
 \Tr_{\sigma} \Big[  (g_R-g_A) \, \, \jh_a   
 \,(g_R-g_A)  \,  \jh_b    \Big] 
\;  \Tr[ \nabla_a Q\, \nabla_b Q ] \cr
&& \equiv \frac{- S_{j k}}{8}   \; 
 \Tr[ \nabla_j Q\, \nabla_k Q ]  
 \label{longitudinal}
\end{eqnarray}
where $\tdj \equiv   \jh_x \frac{\pd_x}{i} + \jh_y \frac{\pd_y}{i}$, 
$\jh_x = \sigma_x , \jh_y = - \sigma_y  $.
We identify $S_{j k} = \sigma^{\rm cf}_{xx} \, \delta_{j k} $ as the dc longitudinal conductivity:
\begin{eqnarray}
\sigma^{\rm cf}_{xx} 
& = &
- \frac{1}{2}
 \Tr  \Big[  (g_R-g_A) \, \, \sigma_x   
 \,(g_R-g_A)  \,  \sigma_x    \Big]   \\
& = &
1+
\frac{ (4 m_+^2 -4 m_-^2 -\alpha_1^2+ \alpha_2^2) 
( \pi + 2\rm{arccot}
[\frac{| m_+ \, \alpha_2 + m_- \, \alpha_1|}
{4 m_+^2 -4 m_-^2 + \alpha_1^2 -  \alpha_2^2}  ])  }
{ 8 | m_+ \, \alpha_2 + m_- \, \alpha_1|} \cr
& \approx &
\frac{ \pi (m^2_+  -  m_-^2 )}{  | m_+ \, \alpha_2 + m_- \, \alpha_1 | } \gg 1.
\end{eqnarray}
The last expression uses the weak disorder limit, i.e.,
$\mathcal{O}(E_F) \approx \mathcal{O}(m) \gg  \alpha_2,\alpha_1$.

Here and below we use the dc ($\omega \rightarrow 0$) limit of the Kubo formula:
\begin{eqnarray}
\sigma^{\rm cf}_{ij}  ( q \to 0 , \omega+ i \eta)  
& = & 
- \int \frac{ d^2 \bm{r'}  }{V}
\int d^2\bm{r} 
\,  
\frac{1}{2 \pi \omega}  \int_{-\infty}^{\infty} dz 
  \nn \\
& \times &  \Bigg(   
 \int_{-\infty}^{\infty} dz 
 \Bigg(
[f(z) -f(z-\omega) ]\; 
\Tr[ \jh_i G_z^R(r,r') \jh_j \,G_{z-\omega}^A(r',r) ]   \nn \\
& + & f(z)\; \Tr[ \jh_i G_{z+ \omega}^R(r,r') \jh_j \,G_{z}^R(r',r) ] \cr
&- &  f(z)\; \Tr[ \jh_i G_{z}^A(r,r') \jh_j \,G_{z-\omega}^A(r',r) ] 
\Bigg).
\label{Kubo}
\end{eqnarray}
For the Dirac theory, the SCBA Green's function is
\begin{eqnarray}
G_D  (\epsilon +  E_F    ; \omega )_{R/A} = 
\frac{1}{  (\epsilon + E_F  + \omega) \sigma_0
- m_D \sigma_z 
+ i \pd_x \sigma_x -i \pd_y \sigma_y 
\pm 
 i \frac{ \alpha_1 \sigma_z + \alpha_2 \sigma_0  }{2}  
 }.
\end{eqnarray}
For the linearized HLR theory, 
\begin{eqnarray}
g(\epsilon+ E_F  ;  \omega )_{R/A} = 
\frac{1}{ 
   (\epsilon+ E_F)  \sigma_0
 +   
  \omega \, P_1  + 2 \Eb \,  P_2  
 + i \pd_x \sigma_x -i \pd_y \sigma_y 
\pm  
i \frac{ \alpha_1 \sigma_z + \alpha_2 \sigma_0  }{2}
 },
\end{eqnarray}
where $P_1 = \frac{1 + \sigma_z }{2} , \; P_2 = \frac{1 - \sigma_z }{2}$. 
The frequency term should be thought as ``mass" term in the Hamiltonian instead of the physical frequency appear in \eqref{Kubo}; 
 otherwise it's unable to match the object defined in Kubo formula. 
Notice that since we've included the Fermi level in the above Greens function, we only need to perform the energy integral $\int dz f(z)$ up to zero. 
We comment that a direct calculation of the Hall conductivity using the above SCBA Green's functions agrees with the results below, up to the crucial additive term equal to ${1 \over 2}$.

\subsection{Goldstone Parameterization}
\label{axialmass}

In the main text we argued that the massless Goldstone modes correspond to $Q_1 = Q_2$ fluctuations, where the matrix bosons $Q_1, Q_2 \in U(2n)/U(n) \times U(n)$.
The $Q_1 = Q_2$ fluctuations correspond to ``vector" gauge transformations of $\Psi$. 
Here we show explicitly that ``axial" gauge transformations corresponding to $Q_1 \neq Q_2$ fluctuations are massive.

Consider the two-field sigma model
  \begin{eqnarray}
&&
S[Q_1 , Q_2 ]
 \equiv 
\Tr ln\Big[  
\Big(  
  \Ea  \, \sigma_0
  - \Eb \, \sigma_z
+ i \sigma_x \pd_x - i \sigma_y \pd_y     \Big) 
+   
\begin{pmatrix}
i \Gamma_1 \, Q_1  & 0 \\
  0     &  i \Gamma_2 Q_2
\end{pmatrix}_\sigma   \;\;  	   \Big] 
\qquad 
\end{eqnarray}
Using the same logic that produced \eqref{longitudinal} with $\hat{H} = (\frac{ \alpha_1 \sigma_z  
+  \alpha_2  \bm{1}_\sigma   }{2}   ) = i \Gamma_1 \pju + i\Gamma_2 \pjd,
 ~ \pju=\frac{1+ \sigma_z}{2},  \pjd= \frac{1- \sigma_z}{2}$, and including a test function $f$,  we calculate
\begin{eqnarray}
&& 
 (g^{-1}_+   - i \, \hat{H}   \tau^3
  + i  \Gamma_1 \,Q_1 \pju + i \Gamma_2 Q_2 \pjd ) \, (\hat{H}  )^{-1}
  \,   
  (g^{-1}_-   +   i \, \hat{H} \, \tau^3 
      -  i \Gamma_1 Q_1 \pju -i Q_2 \pjd  )  \; f    \qquad  \nn \\
&&
=
   g^{-1}_+  \,  \hat{H}^{-1}\,    g^{-1}_-  \, f
  + \ds ( Q_V  \bm{1}_\sigma   +  Q_A \, \sigma_z )\, f 
 + (\ds f) \,(Q_V  \bm{1}_\sigma   +  Q_A \, \sigma_z  ) 
     -  ( Q_V  \bm{1}_\sigma   +  Q_A \, \sigma_z  ) \, (\ds f)       \nn \\
&&
=
    g^{-1}_+  \,   \hat{H}^{-1} \,    g^{-1}_-  \, f
  + \ds ( Q_V  \bm{1}_\sigma   +  Q_A \, \sigma_z )\, f 
 - 2 Q_A \, \sigma_z \, (\ds f),      
\end{eqnarray}
where we've used $Q_1^2 = Q_2^2 = \bm{1}$ and introduced ``vector" $Q_V = {1 \over 2}(Q_1 + Q_2)$ and ``axial" $Q_A = {1 \over 2} (Q_1 - Q_2)$. 
The crucial term that leads to a mass for $Q_A$ is $ - 2 Q_A \, \sigma_z \, \ds $.
We find
  \begin{eqnarray}
&&
{\rm Re} [S[Q_1 , Q_2 ]  \, ]
=
 \frac{1}{2} \Tr \ln \Big[
(g_+^{-1}  \hat{H}^{-1} g_-^{-1})
\Big(
\bm{1}
+ ( g_- \,  \hat{H}  \,   g_+   )
\,(  \ds Q_V  + \ds Q_A \sigma_z   - 2 Q_A \sigma_z \ds  )
\Big)    
\Big]   \nn] \\
&&
=
\frac{1}{16}
\Tr\Big[    (g_R-g_A) 
\Big(  \ds Q_V + \ds Q_A \sigma_z -2 Q_A \sigma_z \ds  \Big)
 \, (g_R-g_A)
  \Big(  \ds Q_V + \ds Q_A \sigma_z -2 Q_A \sigma_z \ds  \Big)  \;   \Big] \cr
 \label{Re-QVQA}
\end{eqnarray} 
Consider the last term quadratic in $Q_A \ds$:
\begin{eqnarray}
&& 
\frac{1}{4}
\Tr\Big[    (g_R-g_A) \Big(    Q_A \sigma_z \ds  \Big)
 \, (g_R-g_A) \Big(    Q_A \sigma_z \ds  \Big)  \;   \Big] \\
 && = \int_{x,x_1,x_2,x_3}
 \int_{k_1 k_2 k_3 k_4}
\frac{1}{4}
\Tr \Big[   
\langle x | (g_R-g_A)    | k_1 \rangle \langle k_1  | x_1 \rangle
\langle x_1 |   Q_A  \sigma_z 
    \ds   | k_2 \rangle \langle k_2   | x_2 \rangle  \nn \\
&&  \qquad \qquad 
 \langle x_2 |  (g_R-g_A)  | k_3 \rangle  \langle k_3   | x_3 \rangle
 \langle x_3 |  Q_A \sigma_z   \ds 
                |  k_4 \rangle  \langle k_4    |  x \rangle 
   \;   \Big]   \\
&&
=
\frac{-1}{4}
\int_{k_1,k_2}
\Tr[  
 (g_R-g_A)\Big{|}_{k_1} \, Q_A(k_2- k_1) \sigma_z \slashed{k_2}
 (g_R-g_A)\Big{|}_{k_2} \, Q_A(k_1- k_2) \sigma_z  \slashed{k_1}
 ]    \\
&&
=  \frac{-1}{4}
\int_p
Q_A(p) Q_A(-p) 
\int_{k_1}
\Tr[ (g_R-g_A)\big{|}_{k_1} \sigma_z (\slashed{p}+\slashed{k_1} ) 
 (g_R-g_A)\big{|}_{k_1+p} \, \sigma_z \slashed{k_1} ]  \\
&&   
\equiv \frac{-1}{4}
\int_p
Q_A(p) Q_A(-p)  \, F(p)   \nn \\
&&
= 
\frac{-1}{4}
\int_p
Q_A(p) Q_A(-p)  \, 
\Big(  F(p=0) + \frac{\pd F}{\pd \slashed{p}  } \Big{|}_{p=0} \, \slashed{p}
+\frac{1}{2} \frac{\pd^2 F}{\pd \slashed{p} ^2} \Big{|}_{p=0} \, \slashed{p}^2
+ ...
\Big),
\label{QAQA-1}
\end{eqnarray}
where we made the change of variables $ k_2 - k_1 \equiv p $. 

Next, examine the two crossing terms containing  $Q_A \ds$ and 
$\ds Q_A $:
\begin{eqnarray}
&&
\frac{1}{16}
\Tr\Big[    (g_R-g_A) 
(  \ds Q_A \sigma_z  ) 
 \, (g_R-g_A) 
(-2 Q_A \sigma_z \ds) 
+ 
(g_R-g_A) 
(-2 Q_A \sigma_z \ds) 
 \, (g_R-g_A) 
 (\ds Q_A \sigma_z )
 \;   \Big]   \nn \\
&& 
=
\frac{-1}{4}
\Tr\Big[    (g_R-g_A) 
(  \ds Q_A \sigma_z  ) 
 \, (g_R-g_A) 
( Q_A \sigma_z \ds) \Big]   \\
&&
=
\frac{-1}{4} \int_{ dp}
Q_A(p) \,Q_A(-p) 
\int_{dk_1}
\Tr\Big[
(g_R-g_A)\Big{|}_{k_1}
{\color{black}\,\sigma_z 
\, \slashed{k_1} 
}
(g_R-g_A)\Big{|}_{p+k_1}
\sigma_z \slashed{k_1}
\Big] 
\label{QAQA-2}
\end{eqnarray}
Combining Eq.~\eqref{QAQA-1} and \eqref{QAQA-2}, the total mass term for $Q_A$ is $\frac{-1}{2}\int_{dp_1}Q_A(p_1) Q_A(-p_1) F(0)$.
The function $F(0) \neq 0$ generally: $Q_A$ is only massless if it's tuned to criticality.
This is in sharp contrast to $Q_V$, which is massless because it's a Goldstone boson.
Thus, $Q_A$ is generally massive and we neglect it at low energies.

\subsection{Imaginary Part of $S_{\rm NLSM}$}

To calculate the imaginary part of $S_{\rm NLSM}$ in \eqref{appendixnlsm}, we first set $n = 1$ and perform the gauge transformation
$\Psi({\bf x}) \rightarrow U({\bf x}) \Psi({\bf x})$ to introduce the 
gauge field $A_j({\bf x}) = i U^\dagger({\bf x}) \partial_j U ({\bf x}) 
 = \sum_{b=1}^3  A_j^b \, \tau_b  
$ where $(\tau_1, \tau_2, \tau_3)$ are the Pauli matrices in retarded-advanced space. 
We determine ${\rm Im}\big(S_{\rm NLSM} \big)$ from the imaginary part of the expectation value of the current (calculated to linear order in $A_j$): 
\begin{align}
i {\delta \big( {\rm Im}S_{\rm NLSM} \big) \over \delta A_j^b} = {\delta S_{\rm top}^{I} \over \delta A_j^b} + {\delta S_{\rm top}^{II} \over \delta A_j^b},
\end{align}
where $S_{\rm top}^{I}$ and $S_{\rm top}^{II}$ are the linear (``quantum") and quadratic (``classical") in $A_j$ contributions to the imaginary part of $S_{\rm NLSM}$ \cite{Streda_1982}.

\subsubsection{$S_{\rm top}^{II}$}
\label{toptwo}

We begin with the ``classical" contribution $S_{\rm top}^{II}$.
By direct evaluation we find
\begin{eqnarray}
S_{\rm top}^{II}  & = &  
 \frac{1}{2}
\Tr[ \jh_\mu \, A_\mu  \, g_+  \jh_\nu   \, A_\nu  \,  g_+  \;  
-    
 \jh_\mu \, A_\mu  \, g_-  \jh_\nu   \, A_\nu  \,    g_-     ]  \cr     
& = &
  \frac{1}{2}
  \Tr  [ \jh_\mu \, A_\mu ( g_+  + g_-   )  \; \jh_\nu \, A_\nu ( g_+  - g_-   )   ]      \nn \cr
& = &   
  \frac{i}{2}
  \Tr_\sigma  [ \jh_\mu \,  ( g_R  + g_A   )  \; \jh_\nu \, ( g_R  - g_A   )   ]
  \, \Tr_{\tau} [  A_\mu   A_\nu  \, \tau_3 ] \cr
& = &   
  {i \over 4} \Tr_\sigma  [ \jh_x \,  ( g_R  + g_A   )  \jh_y \, ( g_R  - g_A   )   ] 
   \Tr_{\tau} [Q (\pd_x Q) (\pd_y Q)]  \nn \\ 
& \equiv &   
i  \frac{\theta^{II} }{2 \pi} 
   \Tr_{\tau} [Q (\pd_x Q) (\pd_y Q)] 
   \label{sigmaXYI}
\end{eqnarray}
where we used $\Tr_{\tau} [Q (\pd_x Q) (\pd_y Q)] = 4 i (A^1_i A^2_j - A_i^2  A_j^1) \epsilon_{ij }$.
${\theta^{II} \over 2\pi}$ is the ``classical" contribution $\sigma_{xy}^{I}$ to the dc Hall conductivity $\sigma^{\rm cf}_{xy} = \sigma_{xy}^I + \sigma_{xy}^{II}$:
\begin{eqnarray}
\sigma^I_{xy} & = & 
\frac{1}{2\pi}
\Big( 
\frac{  2m(\alpha_1+\alpha_2 )  - \mu_F (\alpha_2-\alpha_1) }
{  2m(\alpha_1+\alpha_2 )  + \mu_F (\alpha_2-\alpha_1)  }
\Big) 
\Big(
\tan^{-1}[\frac{  4m   }{\alpha_2 -\alpha_1}]
+ 
\tan^{-1}[\frac{  2 E_F   }{\alpha_2 +\alpha_1}]  
\Big)   \nn \\
&
= & \frac{-1}{2\pi}
\Big( 
\frac{    m_- \Gamma_+  -  m_+ \Gamma_-}
{  m_+ \Gamma_+  - m_-  \Gamma_-   }
\Big) 
\Big(
\arctan [ \frac{2 (\Gamma_+ + \Gamma_- )}{ 2(m_+  + m_-) }]
+
\arctan  [ \frac{2 (\Gamma_+ - \Gamma_- )}{ 2(m_+  - m_-) }]
\Big).
\end{eqnarray}
The ``quantum" contribution $\sigma_{xy}^{II}$ to the Hall conductivity equals ${\theta^{I} \over 2\pi}$ (modulo 1); it's calculated in the next section.

\subsubsection{$S_{\rm top}^{I}$}
\label{topone}

$S_{\rm top}^{I}$ obtains from a result first obtained by Goldstone and Wilczek \cite{PhysRevLett.47.986}; below we follow the treatment in \cite{PhysRevB.77.235431}.
We first introduce the ``mass field,"
\begin{align}
\label{fixedpoint}
\Phi_m(0) \equiv    
 \Ea \sigma_0
   - \Eb \sigma_z  
  + i (\frac{ \alpha_1 \sigma_z + \alpha_2 \sigma_0  }{2})  
\, \tau_3. 
\end{align}
For this calculation, we'll treat $\Phi_m(0)$ as a spatially-varying field $\Phi_m({\bf x})$ that takes its fixed point value \eqref{fixedpoint} at the end of the calculation.
We parameterize the ``mass field" as 
\begin{align}
\Phi_m(x) & \equiv m_1(x) \sigma_z + m_2(x) \sigma_0 \cr
& \equiv \big( m_{1a}(x) \, \tau_3+ m_{1b}(x) \, \tau_0 \big)  \,   \sigma_z 
+ \big( m_{2a}(x) \, \tau_3+ m_{2 b }(x) \, \tau_0 \big) \sigma_0 .
\end{align}
The fixed point values of these masses in two theories are
\begin{eqnarray}
\begin{cases}
(\text{HLR})\quad
m_1  =  -\Eb  + i \frac{\alpha_1}{2}   \tau_z \;,\;\;
m_2 =   \Ea  + i \frac{\alpha_2}{2}  \tau_z
  \\
( \text{Dirac})\;\;   
m_1 = -m_D +  i \frac{\alpha_1}{2}   \tau_z  , \;\;
m_2 = E_F  +   i \frac{\alpha_2}{2}   \tau_z   \\
\end{cases}.
\end{eqnarray}
The real space Green's function is then expanded about uniform $\Phi_m$ as
\begin{eqnarray}
S(\Phi_m) & =& \int \frac{dp_1 dp_2}{(2\pi)^2}
\frac{1}{i \jh_\mu D_\mu + \Phi_m(x ) } e^{i \bm{p} \cdot (x-y)} \cr
& = & S_0
 +
 \Big(  -  S_0    \big(  x_\nu \pd_\nu \Phi_m(0)   \big) \;   S_0  \,   \Big)
 + \ldots,
\end{eqnarray}
where we drop the $``..."$ in what follows.
$S_0$ is the Green's function for uniform $\Phi_m $ with $S_0 \Big{|}_{A_j =0} = g_+$, defined \eqref{gpm-def}

The ``quantum" current ${\delta S_{\rm top}^{I} \over \delta A_j^b}$ is then computed as
\begin{eqnarray}
{\delta S_{\rm top}^{I} \over \delta A_j^b} & = & - \lim_{y \to x} \Tr [ \langle  \jh_j  \, \tau_b
 \,   S(x,y) \rangle ] \cr
 & \approx &
\langle x | 
\Tr [  \jh_j  \, \tau_b \, S_0 \, 
\pd_k \Phi_m(0)\,
\, i S_0 \,\jh_k \,S_0   \; ]
| x \rangle ,
\end{eqnarray}
where we use 
$\langle x' | S_0 \,  \hat{r}_k | x \rangle 
= \langle x' | i S_0 \,\jh_k \,S_0 | x \rangle$.
We set the gauge potential in the Green's function $S_0$ to zero to find
\begin{align}
{\delta S_{\rm top}^{I} \over \delta A_j^b} & = \int \frac{dq_x dq_y}{(2\pi)^2}
\Tr  [ \jh_j\, \tau_b
\frac{1}{  \slashed{q} + \Phi_m(0) }
\pd_k \Phi_m(0)\,
\frac{1}{  \slashed{q} + \Phi_m(0) }
i\, \jh_k 
\frac{1}{  \slashed{q} + \Phi_m(0)   }
] \cr
& =
\frac{-1}{2\pi} \epsilon_{j k}\; \delta_{b 3}
\Big[
\frac{  (m_{2a}+m_{2b})  \, \pd_k (m_{1a}+m_{1b}) 
- (m_{1a}+m_{1b}) \, \pd_k (m_{2a}+m_{2b})    }
{(m_{1a}+m_{1b})^2 - (m_{2a}+m_{2b})^2}  \cr
& -
\frac{  ( - m_{2a}+m_{2b})  \, \pd_k (-m_{1a}+m_{1b}) 
- (-m_{1a}+m_{1b}) \, \pd_k (-m_{2a}+m_{2b})    }
{(-m_{1a}+m_{1b})^2 - (-m_{2a}+m_{2b})^2}
\; 
\Big] \cr
\equiv
\frac{i}{2\pi} \delta_{b 3} \epsilon_{j k} \pd_k \,\theta^{I},
\label{GW-current}
\end{align}
where $\slashed{q} \equiv  \jh_x p_x + \jh_y p_y$.
We now deduce $S^I_{\rm top}$ by coupling this current to $A_j^b$ and then performing an integration by parts:
\begin{eqnarray}
S^{I}_{\rm top} & = & \int d^2x  \;
    \sum_{b = 1}^3   A^b_j \; 
 {\delta S_{\rm top}^{I} \over \delta A_j^b} \cr
& = & { i\over 2 \pi} \int d^2 x \;   A^3_j \; 
\epsilon_{j k}  \pd_k \theta^I \cr
& = &
- {1 \over 4 \pi} \int d^2 x \;   
\theta^I \epsilon_{j k} 
\Tr_{\tau}[\tau^3 \pd_k \; A_j   ].
\end{eqnarray}
 
We now explain in detail how to obtain the specific value of $\theta^I$ for the fixed point value $\Phi_m(0)$ \eqref{fixedpoint}.
This complements the discussion in the main text.
To translate to the notation in the main text, we replace
$m_2 \to \varphi_1 , m_1\to -\varphi_2  $, 
$  \frac{\alpha_2 + \alpha_1}{2}  =\Gamma_1  $,
$ \frac{\alpha_2 - \alpha_1}{2} = \Gamma_2 $.
We also use subscript $s = R,A$ to label the retarded and advanced components in $\tau$ space. 

In Eq.~\ref{GW-current}, we observe that the retarded and advanced contributions are fully separable, i.e., $\theta^I$ only couples to ${1 \over 2} {\rm Tr}(\tau^3 A_j) = A_j^3$.
This allows us to determine $\theta^I$ as the contribution from the $U(1)$ subgroup of $SU(2)$.
We write
\begin{eqnarray}
\Phi^s_m (x ) = -\varphi^s_2(x)   \sigma_z +\varphi^s_1(x) \sigma_0
\equiv 
\begin{pmatrix}
- e^{  - i \chi^s_2 } & 0 \\
0 & e^{i \chi_1^s}
\end{pmatrix}.
\end{eqnarray} 
Plugging into Eq.~\ref{GW-current}, we find
\begin{eqnarray}
- \pd_k \theta^I = \partial_k {\chi_1^R  + \chi_2^R - \chi_1^A - \chi_2^A \over 2}.
\label{Theta-GW-1}
\end{eqnarray}
In the above parameterizations, $\chi_1, \chi_2$ are generally complex. 
We restrict that ${\rm Re}[\chi_{1,2}] \in [0, \pi)$.
We use $ \varphi_1 + \varphi_2 = e^{i \chi_1},  \; \varphi_1 -\varphi_2 = - e^{-i \chi_2 }$ and \eqref{fixedpoint} to find the equations:
\begin{eqnarray}
&&
2m + i \Gamma_2 \,   =  e^{i \chi^R_1}   , \qquad
E_F   + i \Gamma_1   = - e^{-i \chi^R_2},   \\
&&
2m - i \Gamma_2 \,   =  e^{i \chi^A_1}   , \qquad
E_F   - i \Gamma_1   = - e^{-i \chi^A_2}.
\end{eqnarray}
For retarded field $\chi_{1,2}^R$, we find the solution:
\begin{eqnarray}
&&   
{\rm Re}[\chi^R_1] = \arctan[\frac{ \Gamma_2 }{2m}] , \;\;
{\rm Im}[\chi^R_1] =  \frac{-1}{2} \log[(2m)^2 + (\Gamma_2)^2],  \\
&&
{\rm Re}[\chi^R_2] = \pi- \arctan[\frac{ \Gamma_1 }{E_F}] , \;\;
{\rm Im}[\chi^R_2] =  \frac{-1}{2} \log[(E_F)^2 + (\Gamma_1)^2].
\end{eqnarray}
Due to the restriction ${\rm Re}[\chi_{1,2}] \in [0, \pi)$, we need to perform 
a charge conjugation on the advanced fermion part of the action $S_A \to {\cal C} S_A {\cal C} $ defined in 
Eq.~\ref{chargeeuclidean}.
It is in this step that we use the fact that $\theta^I$ is only sensitive to the Abelian ${1 \over 2} {\rm Tr} (\tau^3 A_j)$ component of $A_j$.
Charge conjugation flips the sign of mass term involving $\sigma_0$ as well as the gauge coupling of the advanced fermion to $A_j^3$:
\begin{eqnarray}
&&  
S_A =\Psi^\dagger_A
\Big( 
  \Ea   \sigma_0
   - \Eb \sigma_z 
   +
 i \sigma_x (\partial_x - i A^3_x )
 - i \sigma_y  (\partial_y - i A^3_y )
   - i \frac{ \alpha_1 \sigma_z + \alpha_2 \sigma_0  }{2}
\Big)  \; \Psi_A,   \\
&&  
 {\cal C} S_A {\cal C}   =\Psi^\dagger_A
\Big( 
  - \Ea   \sigma_0
   - \Eb \sigma_z 
   +
 i \sigma_x (\partial_x + i A^3_x )
 - i \sigma_y  (\partial_y + i A^3_y )
   - i \frac{ \alpha_1 \sigma_z - \alpha_2 \sigma_0  }{2}
\Big)   \Psi_A. \qquad  
\end{eqnarray}

In this new basis, i.e., after the charge conjugation, we find the current ${\delta S_{\rm top}^{I} \over \delta A_j^b}$ equals
\begin{eqnarray}
&&
{\delta S_{\rm top}^{I} \over \delta A_j^b}
=
\frac{-1}{2\pi} \epsilon_{j k}\; \delta_{b 3}
\Big[
\frac{  (m_{2a}+m_{2b})  \, \pd_k (m_{1a}+m_{1b}) 
- (m_{1a}+m_{1b}) \, \pd_k (m_{2a}+m_{2b})    }
{(m_{1a}+m_{1b})^2 - (m_{2a}+m_{2b})^2}  \nn \\
&& \qquad \qquad  
+ 
\frac{  (-1) \,  ( - m_{2a}+m_{2b})  \, \pd_k (-m_{1a}+m_{1b}) 
- (-m_{1a}+m_{1b}) \,(-1)  \pd_k (-m_{2a}+m_{2b})    }
{(-m_{1a}+m_{1b})^2 - (-m_{2a}+m_{2b})^2}
\; 
\Big]  \nn  \\
&&
\equiv
\frac{i}{2\pi} \delta_{b 3} \epsilon_{j k}
\pd_k \,\theta^I.
\end{eqnarray} 
The retarded contribution to $\theta^I$ is unchanged.
The advanced fermion mass becomes
\begin{eqnarray}
&&
\Phi^A_m (x ) = -\varphi^A_2(x)   \sigma_z + (-1)\varphi^A_1(x) \sigma_0
\equiv 
\begin{pmatrix}
- e^{  - i \tilde{\chi}^A_2    } & 0 \\
0 & e^{i \tilde{\chi}_1^A }
\end{pmatrix}
\end{eqnarray} 
where $  \phi^A_2  = - m^A_1 =  \Eb  + \frac{i \alpha_1 }{2}, \;\;
\varphi^A_1(x)  =  m_2^A  = \Ea - \frac{i \alpha_2}{2} $.
This results in the equations:
\begin{eqnarray}
&&
2m - \Gamma_2  = e^{i \tilde{\chi}^A_2} , \qquad
-E_F  +  i\Gamma_1   = e^{i \tilde{\chi}^A_1},   \\
&&
{\rm Re}[\tilde{\chi}^A_2 ]= \arctan[\frac{\Gamma_2}{2m} ] , \qquad
{\rm Re}[\tilde{\chi}_1^A]  =  \pi -\arctan[\frac{\Gamma_1}{E_F}]
\end{eqnarray}
Since charge conjugation also flips relative sign between retarded and advanced contributions in Eq.~\ref{Theta-GW-1}, we have
\begin{eqnarray}
 \theta^I & = & - \frac{ 1}{2} \Big[  
 ({\rm Re}[ \chi_1^R]  + {\rm Re}[\chi_2^R ]+ {\rm Re}[\tilde{\chi}_1^A ]
 + {\rm Re}[\tilde{\chi}_2^A] )   
\Big] \\
& = & \pi + \arctan[\frac{\Gamma_+ - \Gamma_-}{ m_{+}- m_{-}   }] - \arctan[\frac{ \Gamma_+ + \Gamma_- }{  m_{+}+ m_{-}  }]. 
\label{sigmaXYII}
\end{eqnarray}
We have used $\theta^I = 0$ (mod $2\pi$) to fix the coefficient of $\pi$ to be unity. 

\bibliography{bigbib}

\end{document}